# PROJECT X FUNCTIONAL REQUIREMENTS SPECIFICATION *

S. D. Holmes, S. D. Henderson, R. Kephart, J. Kerby, I. Kourbanis, V. Lebedev, S. Mishra, S. Nagaitsev, N. Solyak, R. Tschirhart, Fermilab, Batavia, IL 60510, USA

*Abstract*

Project X is a multi-megawatt proton facility being developed to support a world-leading program in Intensity Frontier physics at Fermilab. The facility is designed to support programs in elementary particle and nuclear physics, with possible applications to nuclear energy research. A Functional Requirements Specification has been developed in order to establish performance criteria for the Project X complex in support of these multiple missions, and to assure that the facility is designed with sufficient upgrade capability to provide U.S. leadership for many decades to come. This paper will briefly review the previously described Functional Requirements, and then discuss their recent evolution.

## PROJECT X MISSION AND GOALS

Project X is a high intensity proton facility that would support a world-leading U.S. program in Intensity Frontier physics over the next several decades. Project X is currently under development by Fermilab with both national and international partners.

The primary mission elements to be supported by Project X include:

1. Provide a neutrino beam for long baseline neutrino oscillation experiments, based on targeting at least 2 MW of proton beam power at any energy between 60 – 120 GeV.

2. Provide MW-class, 1-3 GeV, proton beams supporting multiple kaon, muon, neutrino, and nuclei-based precision experiments. Simultaneous operation with the long baseline neutrino program is required.

3. Provide a path toward a muon source for a possible future Neutrino Factory and/or a Muon Collider.

4. Provide opportunities for nuclear energy applications with high duty factor MW-class beams at 1 GeV.

These elements represent the fundamental design criteria for Project X. A design concept, designated the Reference Design [1], has been established supporting these elements in an innovative and flexible manner. The Reference Design is based on a 3 GeV superconducting CW linac, augmented by a superconducting pulsed linac for acceleration from 3 to 8 GeV.

## PROJECT X REFERENCE DESIGN

The Reference Design has been previously described in [2]. The primary elements are:

- An H- front end consisting of a 5 mA, 30 kV, CW ion source; a low energy beam transport (LEBT) line; a 2.1 MeV CW RFQ; and a medium energy beam transport (MEBT) line containing a wideband chopper capable of accepting or rejecting bunches in arbitrary patterns at up to 162.5 MHz;
- A 3 GeV superconducting linac operating in CW mode and capable of accelerating an average beam current of 1 mA (averaged over >1 μsec), and a peak beam current of 5 mA (averaged over <1 μsec);
- A 3 to 8 GeV pulsed superconducting linac capable of accelerating an average current of 43 μA with a duty factor of 4.3%;
- A pulsed dipole that can split the 3 GeV beam between the pulsed linac and the 3 GeV program;
- An rf splitter that can deliver the 3 GeV beam to at least three experimental areas;
- Experimental facilities to support an initial round of 3 GeV experiments;
- Modifications to the Main Injector and Recycler complex required to support 2 MW operations for the long baseline neutrino program;
- All interconnecting beam lines.

A schematic depiction of the Reference Design is given in Figure 1.

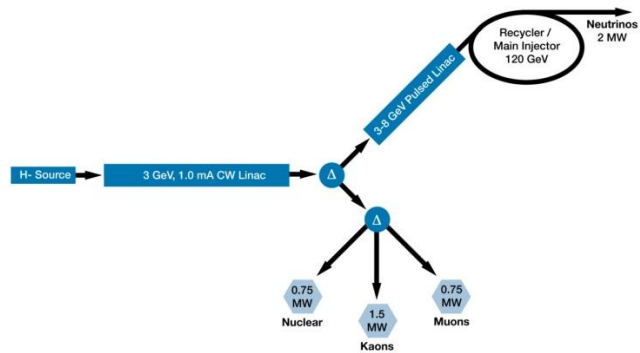

Figure 1: Project X Reference Design

## PROJECT X FUNCTIONAL REQUIREMENTS

A set of Functional Requirements have been established to guide the development of the Reference Design. These have been described in more detail in [3, 4]. The major parameter subset of these is given in Table 1.

* Work supported by the Fermi Research Alliance, under contract to the U.S. Department of Energy

Table 1: Project X Primary Functional Requirements

| Requirement | Description | Value |
|---|---|---|
| **3 GeV Linac** | | |
| L1 | Delivered Beam Energy, maximum | 3 GeV (kinetic) |
| L3 | Average Beam Current (averaged over >1 μsec) | 1 mA |
| L4 | Maximum Beam Current (sustained for <1 μsec) | 5 mA |
| L8 | Minimum Bunch Spacing | 6.2 nsec (1/162.5 MHz) |
| L10 | Bunch Pattern | Programmable |
| L11 | RF Duty Factor | 100% (CW) |
| L12 | RF Frequency | 162.5 MHz and harmonics thereof |
| L13 | 3 GeV Beam Split | Three-way |
| **3-8 GeV Pulsed Linac** | | |
| P1 | Maximum Beam Energy | 8 GeV |
| P3 | Charge to fill Main Injector/cycle | 26 mA-msec in <0.75 sec |
| P5 | Duty Factor (initial) | < 5% |
| **Main Injector/Recycler** | | |
| M1 | Delivered Beam Energy, maximum | 120 GeV |
| M2 | Delivered Beam Energy, minimum | 60 GeV |
| M4 | Beam Power (60-120 GeV) | > 2 MW |
| M7 | Beam Pulse Length | ~10 μsec |
| M11 | Pulse Repetition Rate (120 GeV) | 1.2 sec |
| M12 | Pulse Repetition Rate (60 GeV) | 0.75 sec |
| **Integration** | | |
| I1 | The 3 GeV and neutrino programs must operate simultaneously | |
| **Upgradability** | | |
| U1 | Provisions should be made to support an upgrade of the CW linac to an average current of 5 mA. | |
| U3 | Provisions should be made to deliver CW proton beams as low as 1 GeV. | |
| U5 | Provisions should be made to support an upgrade of the pulsed linac to a duty factor of 10%. | |

The evolution of the functional requirements over the last year has been primarily in the areas of: 1) staging options; 2) upgradability to support muon facilities; and 3) upgradability to support short baseline neutrino experiments.

## *Staging*

Considerations of the total cost of the Reference Design have led to the requirement that Project X be constructible in stages subject to the following criteria applied at each stage:

- The cost to construct each stage should be substantially less than $1 billion;
- Each stage should support a compelling physics research program; and
- The full Reference Design should be realized in the final stage.

A possible set of stages that meets these requirements has been identified as summarized in Table 2.

Table 2: Staging Opportunities for Project X

| | Stage 1 | Stage 2 | Stage 3 |
|---|---|---|---|
| Configuration | 1 GeV CW linac, injecting into existing Booster | Add 1-3 GeV CW linac | Full Reference Design |
| 1 GeV Beam Power | 1 MW | 1 MW | 1 MW |
| 3 GeV Beam Power | --- | 3 MW | 3 MW |
| 8 GeV Beam Power | 115 kW | 115 kW | 350 kW |
| 120 GeV Beam Power | 1.1 MW | 1.1 MW | 2.3 MW |

This particular example provides significant physics research opportunities at each stage, while ultimately realizing the full reference design. The following should be noted in this strategy: 1) maintenance of a 1 MW available beam power at 1 GeV, simultaneous with 3 MW availability at 3 GeV would require an upgrade of the rf system to support 2 mA average current in the first 1 GeV of the linac. A decision to implement such an upgrade as part of Stage 2 would be taken on the basis of programmatic goals following the completion of Stage 1. 2) 8 GeV beams are delivered from the existing Fermilab Booster in Stages 1 and 2, and from the 8 GeV Project X pulsed linac in Stage 3.

### Muon Facilities

The Reference Design configuration provides a powerful source of protons that could provide a platform for a muon-based Neutrino Factory or Muon Collider. However, the Reference Design on its own is insufficient, and will require upgrades to provide the required beam parameters. These upgradability requirements are captured in the "Upgradability" section of the Functional Requirements (Table 1).

The Muon Collider requires approximately 4 MW of beam power delivered onto a production target at an energy between 5 and 15 GeV. The 8 GeV pulsed linac of Project X, suitably modified, can meet this particular need. However, a very low duty factor beam is required, for example the Muon Collider requires beam delivered in a single bunch, with a bunch length of 2-3 nsec, at a 15 Hz repetition rate [5]. Requirements are modestly relaxed for a Neutrino Factory.

Providing the required beam power and bunch structure for a Muon Collider will require both an upgrade of the Project X 8 GeV beam power and additional facilities to reformat the high duty factor beam from Project X. Project X naturally provides 350 kW of beam power at 8 GeV (Table 2). The strategy for providing 4 MW at 15 Hz is to increase the current to 5 mA, increase the pulse length to 6.7 msec, and increase the repetition rate to 15 Hz. The result is a 10% duty factor at 5 mA, or 4 MW delivered from the pulsed linac at 8 GeV. These requirements are documented as requirements U1 and U5 in Table 1.

Additionally the beam delivered at 8 GeV must be reformatted to provide the very low duty factor required for muon facilities. It is believed that two rings will be required: 1) an accumulation ring that collects the 6.7 msec long H- pulse and segregates it into roughly 4-8 bunches; and 2) a compressor ring that reduces the bunch to the required 2-3 nsec. This would be followed by a "trombone" beam line that utilizes varying times of flight to deliver multiple bunches onto the production target simultaneously. These concepts require further development to establish viability.

### Short Baseline Neutrinos

There has been significant recent interest in the development of multi-MW proton sources at low (<10 GeV) energy either for short baseline or low energy long baseline neutrino experiments [6, 7]. These experiments also require low ($10^{-4}$ or less) duty factors. Fortunately these requirements are nearly identical to the Muon Facility requirements. Hence, a common strategy could be employed that would provide both the intermediate term benefit of 4 MW of beam power at 8 GeV for neutrino experiments, but a longer term opportunity for muon-based facility.

### Stage 4

The above considerations have led to the identification of a "Stage 4" configuration of Project X that goes beyond the Reference Design and provides (simultaneously) up to 3 MW at 3 GeV, 4 MW at 8 GeV, and 4 MW at 120 GeV. This configuration is associated with the requirements for the Muon Facilities and Short Baseline Neutrino requirements described above.

## SUMMARY

Project X is central to the strategy for future development of the Fermilab accelerator complex. A Reference Design has been developed that will support a world leading program in neutrinos and other rare processes over the coming decades, and will provide a platform for future muon-based facilities – either a Neutrino Factory or a Muon Collider. A set of Functional Requirements forms the basis of the design, and provides flexibility for staging opportunities and for long-term upgradability of the complex.